\newcommand{\Om}{\Omega}
\newcommand{\ep}{\epsilon}
\newcommand{\eps}{\varepsilon}\newcommand{\de}{\delta}
\newcommand{\vphi}{\varphi}
\newcommand{\intl}{\int\limits}\newcommand{\su}{\sum\limits_{i=1}^2}
\newcommand{\R}{{\Bbb R}}\newcommand{\N}{{\Bbb N}} 
\newcommand{\pa}{\partial}
 
\newcommand{\DD}{{\cal D}} 
\newcommand{\ea}{\end{array}}

\newcommand{\nn}{\nonumber}\newcommand{\ba}{\begin{array}}
\newcommand{\beq}{ \begin{equation} }\newcommand{\eeq}{\end{equation} }
\newcommand{\bea}{\begin{eqnarray}}\newcommand{\eea}{\end{eqnarray}}
\newcommand{\beas}{\begin{eqnarray*}}\newcommand{\eeas}{\end{eqnarray*}}
\newcommand{\beqn}{ \begin{equation*} }

\newcommand{\bblock}{\\ \begin{minipage}{15cm} \beas}
\newcommand{\eblock}[1]{\eeas \end{minipage} 
\hfill \begin{minipage}{10mm}\bea \label{#1}\eea\end{minipage}\\}
\documentstyle[aps,amsfonts]{revtex}
\title{A note on the Penrose junction conditions}
\author{Michael Kunzinger
	\footnote{Electronic mail: Michael.Kunzinger@univie.ac.at}}
\address{Department of Mathematics, University of Vienna, 
	Strudlhofg.~4\\ A-1090 Wien, Austria}
\author{Roland Steinbauer
	\footnote{Electronic mail: roland.steinbauer@univie.ac.at}}
\address{Department of Mathematics, University of Vienna, 
	Strudlhofg.~4\\ Institute for Theoretical Physics, 
	University of Vienna, Boltzmanng.~5\\A-1090 Wien, Austria}

\begin{document}
\maketitle
\thispagestyle{empty}
\begin{abstract}\noindent
Impulsive pp-waves are commonly described either by a distributional 
spacetime metric  or, alternatively,  by  a  continuous  one. 
The transformation $T$ relating these forms 
clearly has to be discontinuous, which causes two basic problems:
First, it changes the manifold structure and second, 
the    pullback  of the distributional form of the metric under
$T$ is not well defined within classical distribution theory.
Nevertheless, from a physical point of view both pictures are equivalent. 
In this work, after calculating $T$ as well as the ``Rosen''-form of 
the metric  in the general case of a pp-wave  with arbitrary wave profile 
we  give  a  precise  meaning  to  the  term ``physicially equivalent'' by 
interpreting $T$ as the distributional 
limit of a suitably regularized  sequence   of diffeomorphisms.  
Moreover,  it is shown that $T$ provides  an  example  of  a
generalized   coordinate   transformation   in   the  sense  of
Colombeau's generalized functions.\\
\nopagebreak

{\em PACS-numbers: }04.20.-q, 04.20.Cv, 04.30.-w, 02.30.Sa, 02.90.+p. 
\begin{flushright}Report no: UWThPh -- 1998 -- 57\end{flushright}
\end{abstract}
\nopagebreak
\section{Introduction}\label{intro}                  
Plane fronted gravitational waves with parallel rays (pp-waves) are spacetimes 
characterized by the existence of a covariantly constant null vector field,
which allows to write the metric in the form~\cite{ksmh}
\beq\label{dm}ds^2\,=\,Hdu^2+2d\zeta d\bar\zeta-2dudv\,,\eeq
where $H(u,\zeta,\bar\zeta)$ is a function of the the retarded time coordinate 
$u$ and the transverse complex coordinates $\zeta$ and $\bar\zeta$.
{\em Impulsive} pp-waves may now be defined by setting~\cite{penrose}
\beq H(u,\zeta,\bar\zeta)\,=\,f(\zeta,\bar\zeta)\,\de(u)\,,\eeq
where $f$ denotes a smooth function of $\zeta,\bar\zeta$ and $\de$ is the
Dirac-$\de$-distribution. Such spacetimes arise in a very natural manner as the
ultrarelativistic limit of Kerr-Newman black holes~\cite{as,ls,bn} and
multipole solutions of the Weyl family as recently shown
by Podolsky and Griffiths~\cite{pg}. Moreover these geometries play an
important role in the description of scattering processes at the Planck 
scale~\cite{vv,deharo}.

While the form (\ref{dm}) of the metric very clearly demonstrates the nature of
the impulsive wave, i.e. that the spacetime is flat everywhere except for the 
null hyperplane $u=0$ where the $\de$-like shock is located,
it has the obvious disadvantage of involving distributional coefficients.
However, due to the simple form of the metric the curvature tensor can be 
calculated and is itself proportional to the $\de$-distribution.
Moreover, the geometry of impulsive pp-waves  was described 
recently~\cite{geo,geo2} by the authors entirely in the distributional 
picture. The key idea was to use a careful regularization procedure 
which allowed to handle the nonlinear singular geodesic and geodesic 
deviation equations in a mathematically satisfactory way.

On the other hand, 
impulsive pp-waves are frequently described by a different spacetime metric
which is actually continuous (see~\cite{penrose,pv1,pv2} and, for the
general case,~\cite{aib}).  
In the special case of an impulsive {\em plane} (linearly polarized) wave,
i.e. $f=1/2(\zeta^2+\bar\zeta^2)$, it takes the form
\beq\label{cm} 
	ds^2\,=\,(1+u_+)^2dX^2+(1-u_+)^2dY^2-2dudV\,.
\eeq
where $u_+$ denotes the ``kink'' function $u_+=\theta(u)\,u$.
This form of the metric has the advantage that only the curvature tensor
involves distributions while the metric can be treated ``classically.''
Moreover, if one constructs impulsive pp-waves by glueing together two copies
of Minkowski space along the hypersurface $u=0$ with a coordinate shift 
(according to Penrose's scissors and paste approach~\cite{penrose}) a theorem
of Clarke and Dray~\cite{cd} ensures the existence of a $C^1$-atlas in
which the metric components are continuous.

Clearly a transformation relating the metrics (\ref{dm}) and (\ref{cm}) cannot
even be continuous, hence strictly speaking it changes (even the topological
structure of) the manifold. In the case of a plane wave this discontinuous 
change of variables is given by the so called Penrose junction 
conditions~\cite{penrose} (for the general vacuum case again see~\cite{pv1})
\bea~\label{jc} 	
	\sqrt{2}\zeta&=&X(1+u_+)+iY(1-u_+)\nn\\
	v&=&V+\frac{1}{2}X^2(u_++\theta(u))+\frac{1}{2}Y^2(u_+-\theta(u))\\
	u&=&u\,.\nn
\eea

However, the two (mathematically) distinct spacetimes are equivalent from a 
physical point of view, i.e. the geodesics and the particle motion 
agree~\cite{bis-proc}. More precisely, if one computes the geodesics for the
continuous metric and then (formally) applies the discontinuous transformation
one obtains exactly the geodesics as derived in the distributional picture
in~\cite{geo}.

Nevertheless,    it    has    to    be  noted  that  the actual
calculation  of  the  continuous  metric  (\ref{cm})  from  the
distributional   form (\ref{dm})  as  well  as  the
transformation  of  the  geodesics  involve undefined nonlinear
operations  with distributions which are dealt with by using ad
hoc  multiplication  rules; in  this  case $\theta^2=\theta$,
$\theta  u_+  =  u_+$,  $u_+  \delta  = 0 =u_+^2\delta$
and $\theta\de=1/2\,\de$. Such a   seemingly 
harmless  operational  approach  to  introducing nonlinearities  into linear
distribution theory may work out in certain  cases  but  in  just  as  many 
instances it will lead seriously  astray. Using the above, i.e.
exactly the ``rules'' appearing in the situation under consideration we get
$(\theta\theta)\de= \theta\de=1/2\,\de\not=1/4\,\de=\theta(\theta\de)$.
For examples of the difficulties that may result from such
manipualtions both in mathematics and in physics, cf. \cite{hajek} and
\cite{stein}.   Especially   in   the   context  of  nonlinear
differential equations involving generalized functions (in this
case: the  geodesic  equations) it turns out that the method of
choice for determining which multiplications work out and which
don't  is to first regularize the singularities, then carry out
the    nonlinear   operations   and,   finally,   compute   the
distributional limits, if they exist (cf. \cite{mo}, p. 180).

By  following  this  route, in this paper  we give a  precise 
meaning to the above-mentioned ``physical equivalence,''
interpreting the discontinuous change of variables as the (distributional) 
limit of a family of smoothened transformations. 
The key fact is to note that the coordinate lines in the new variables defined 
in (\ref{jc}) are exactly the geodesics of the metric (\ref{dm}) with 
vanishing initial speed in the $\zeta-,\bar\zeta-$ and $v-$directions.

More precisely, we start with the distributional form (\ref{dm}) 
of the metric, which we 
turn   into   a   sandwich   wave  by  regularizing  the  $\de$
distribution in a completely general manner. 
Then we carry out a transformation analogous to (\ref{jc})
and calculate the distributional limit of the metric to arrive at
the continuous form, which in the special case considered above 
precisely agrees with (\ref{cm}). 
In calculating this limit, which is totally
independent of the form of the regularization (hence ``natural''), we make 
use of the results in~\cite{geo} and~\cite{geo2}. In the
sandwich wave -- picture both forms of the impulsive wave, i.e. the metrics 
(\ref{dm}) and (\ref{cm}), arise as (distributional) limits in different
coordinate systems (see also the remarks in Sec.~\ref{conclusions}).
\section{The smoothened Transformation}\label{trsf}

Since our considerations depend heavily upon the results of~\cite{geo,geo2}
we shall stick to the real coordinates used there and start with an impulsive 
pp-metric of the form
\beq \label{metric}
ds^2\,=\,f(x^i)\,\de(u)\,du^2-du\,dv+\sum_{i=1}^2(dx^i)^2\,\,, 
\eeq
where $x^i=(x,y)$ denotes the transverse coordinates spanning the wave
surface. The first step is now to regularize the $\de$-distribution by a 
sequence of (smooth) functions $\rho_\eps$ $(0<\eps\leq 1)$. 
Since we are interested in the most general result we will only assume the 
following minimal set of conditions on the regularization (so called 
{\em generalized $\de$-function}\,)
\bea\label{gendelta}
	\mbox{(a)}&\quad&\mbox{\rm supp}(\rho_\eps)\,\subseteq\,
	  [-\eps,\eps]\,\,,\nn\\
	\mbox{(b)}&\quad&\int\rho_\eps(x)\,dx\to 1\quad (\eps\to 0)\,\,
	  \mbox{ and} \\
	\mbox{(c)}&\quad&\int|\rho_\eps (x)|
	   \,dx \le C\mbox{ for small } \eps\,.\nn
\eea
Note that condition (a) is choosen in order to avoid technicalities in the 
calculation of the distributional limits which, however, remain valid if 
(a) is replaced by 
\beq	\mbox{(a')}\qquad\mbox{\rm supp}(\rho_\eps)\to\{0\}\quad(\eps\to 
	0)\,\,.
\eeq
Hence we are left with the following sequence of sandwich waves
\beq\label{sandwich}
	ds_\eps^2\,=\,f(x^i)\,\rho_\eps(u)\,du^2-du\,dv+\sum_{i=1}^2(dx^i)^2\,\,.\eeq
The corresponding geodesic equations are given by\\
\parbox{15cm}{
\beas
	\ddot x^i_\eps(u)&=&\frac{1}{2}\,\pa_i f(x^j_\eps(u))\,\rho_\eps(u)\\
	\ddot v_\eps(u)&=&f(x^j_\eps(u))\,\dot\rho_\eps(u)
              \,+\,2\,\pa_i\,f(x^j_\eps(u))\,\,\dot x^i_\eps(u)\,\rho_\eps(u),
\eeas
} \hfill \parbox{10mm}{\bea \label{georeg}\eea}\\
where we have used $u$ as an affine parameter.
Since we are only interested in geodesics with the special initial conditions
$x^i_\eps(-1)=x^i_0$, $v_\eps(-1)=v_0$ and $\dot x^i_\eps(-1)=0=\dot v_\eps(-1)$, 
we adopt the notation $x^i_\eps(x^i_0,u)$ and $v_\eps(v_0,x^i_0,u)$ for 
these geodesics with vanishing initial speed. They obey the following
(implicit) set of equations \\
\parbox{15cm}{
\beas	
       \,\, x^i_\eps(x^i_0,u)&=&x_0^i+\frac{1}{2}\intl_{-\eps}^u\intl_{-\eps}^s\pa_i
       	  f(x^j_\eps(x_0^k,r))\rho_\eps(r)\,dr\,ds\\
	\,\, v_\eps(v_0,x^i_0,u)&=&v_0+\intl_{-\eps}^u f(x^j_
	\eps(x_0^k,s))
	  \rho_\eps(s)\,ds+
	  \intl_{-\eps}^u\intl_{-\eps}^s\pa_i f(
	  x^j_\eps(x^k_0,r))\dot x^i_\eps(x_0^k,r))\rho_\eps(r)\,dr\,ds
\eeas
} \hfill \parbox{10mm}{\bea \label{geo}\eea}\\
In~\cite{geo2} it was shown that these equations are uniquely solvable (even)
within the framework of Colombeau's algebras of generalized 
functions~\cite{mo,c1,AB}, which provide a suitable setting for singular
differential equations. Moreover the unique solution possesses the following
distributional  limit, {\em independent of the concrete shape of the
regularization}\\
\parbox{15cm}{
\beas
     x^i_\eps(x^j_0,u)&\to&x^i_0+\frac{1}{2}\,\pa_if(x^j_0)\,u_+\nn\\
     v_\eps(v_0,x^j_0,u)&\to&v_0+f(x^j_0)\,\theta(u)+
     	\frac{1}{4}\pa_if(x^j_0)\,\pa^if(x^j_0)\,u_+\,.
\eeas
} \hfill \parbox{10mm}{\bea \label{res1}\eea}\\
Let us now consider the transformation $T_\eps: (u,v,x^i)\to (u,V,X^i)$ 
(where $X^i=(X,Y)$) given implicitly by (\ref{geo}) and 
depending on the regularization parameter $\eps$ according to\\
\parbox{15cm}{
\beas 
	x^i&=&x^i_\eps(X^j,u)\\
	v&=&v_\eps(V,X^j,u)
\eeas
} \hfill \parbox{10mm}{\bea \label{trafo}\eea}\\
which is precisely analogous (and in the limit $\eps\to 0$ and the special
case of a plane wave actually reduces) to (\ref{jc}). 

We  claim that $T_\eps$ yields a coordinate transformation
such that the new coordinates 
$(u,V,X,Y)$ are constant along the geodesics given by (\ref{geo}).
While   the   latter   property   follows   directly  from  
the construction, we still have to
verify that $T_\eps$ is a diffeomorphism 
in  a  spacetime  region  containing the shock-hypersurface
$u=0$. To establish this property we  
employ   a  global  univalence  theorem  by  Gale  and  Nikaido
(\cite{uni},  Thm.  4) stating that any differentiable $F:\Omega
\to  \R^n$,  where  $\Omega$  is a closed rectangular region in
$\R^n$  is univalent (injective) if all principal
minors of its Jacobian $J(x)$  are positive. Since
\beq \label{det}
\frac{\pa(\,u,\,x^1,\,x^2,\,v\,)}{\pa(u,X^1,X^2,V)}
= \left|\matrix{1&0&0&0\cr 
\frac{\pa x^1_\eps}{\pa u}& \frac{\pa x^1_\eps}{\pa X^1}& 
\frac{\pa x^1_\eps}{\pa X^2} & 0\cr 
\frac{\pa x^2_\eps}{\pa u}& \frac{\pa x^2_\eps}{\pa X^1}& 
\frac{\pa x^2_\eps}{\pa X^2} & 0\cr 
\frac{\pa v_\eps}{\pa u}& \frac{\pa v_\eps}{\pa X^1}& 
\frac{\pa v_\eps}{\pa X^2} & 1\cr }\right|
\eeq
we have to find estimates for
\beq \label{dxX}
\frac{\pa x^i_\eps}{\pa X^j} = \delta^i_j + 
\frac{1}{2}\intl_{-\eps}^u\intl_{-\eps}^s(\pa_m\pa_i f)
(x^k_\eps(X^l,r)) \frac{\pa x^m_\eps}{\pa X^j}(X^l,r)
\rho_\eps(r)\,dr\,ds\,.
\eeq
If $X^l$ varies in a compact region $K$ of $\R^2$
and $-1 \le p \le u < \infty$ it follows from a straightforward
modification  of  the  appendix  of  \cite{geo} that for small 
$\eps$ the terms $x^k_\eps(X^l,p)$  remain  bounded,  independently  
of   $X^l$,  $p$  and  $\eps$.  
Let $g(u):=\sup\{\su |\frac{\pa x^i_\eps}{\pa X^j}(X^k,p)|\,:\,
X^k \in K, \, -1 \le p \le u\}$. Then (\ref{dxX}) gives
\[
|g(u)| \le C_1 + C_2 \intl_{-\eps}^u |g(s)|\,ds\, ,
\]
so Gronwall's  lemma implies the same boundedness property for 
$\frac{\pa   x^i_\eps}{\pa  X^j}$.  Using  these  estimates  it
follows  from  (\ref{dxX})  that for small $\eps$ and for small
$u>0$  (depending on $\|\pa_i\pa_j f\|_\infty$ in a compact region and
on $C$ from (\ref{gendelta}c)) $\frac{\pa x^i_\eps}{\pa X^j}$
will  remain  arbitrarily  small (for $i\not=j$) or arbitrarily
close  to  $1$  (for  $i=j$), respectively. Hence all principal
minors  of  (\ref{det})  are indeed positive in a suitable rectangular
region (independent of $\ep$)  containing   the   shock   hypersurface   
$u=0$  which
establishes  our  claim. We note that this result in particular
implies  that for small $\eps$ the geodesics (\ref{geo}) do not
cross in this region.

In the new coordinates $(u,V,X,Y)$ the metric tensor takes the 
form
\beq\label{trsfm}
     ds_\eps^2\,=\,
	-dudV+(2\sum\limits_{i=1}^2(\dot x^i_\eps\pa_jx^i_\eps)-\pa_jv_\eps)
	dudX^j
	+\sum\limits_{i=1}^2(\pa_jx^i_\eps dX^j)^2\,,
\eeq
where
$\dot{\,}$ and $\pa_i$ denote derivatives with respect to $u$ and $X^i$ 
respectively and we  have  omitted  the  straightforward
computation showing that $g_{uu}=0$.

\section{Distributional Limit}
Our next task is to compute the distributional limit of the metric 
(\ref{trsfm})   as   $\eps\to  0$.  To  reduce  the  notational
complexity  we are going to suppress the explicit dependence of
test    functions  on  the  variables  $X^i$  and  $V$, i.e. we
will  write $\vphi(u)$ for $\vphi\in \DD(\R^4)$ and will simply
drop the integrations with respect to $X^i$  and  $V$.
This abuse of notation is 
admissible   due   to   the   uniform   boundedness  properties
(in $X^i$ and $V$) established  above and serves to reduce 
the number of integrals in the sequel by $3$.
We start out with the coefficient\\
\parbox{15cm}{
\beas 
 g_{uX}&=&\su(\intl_{-\eps}^u\pa_i f(x^j_\eps(X^k,s))\rho_\eps(s)\,ds
 \,\,\pa_1(X^i+\frac{1}{2}\intl_{-\eps}^u\intl_{-\eps}^s
 \pa_i f(x^j_\eps(X^k,r)\rho_\eps(r)\,drds))\\
  &&-\pa_1(V+\intl_{-\eps}^uf(x^j_\eps(X^k,r))\rho_{\eps}\,dr
 +\intl_{-\eps}^u\intl_{-\eps}^s\pa_lf(x^j_\eps(X^k,r)
 \dot x^l_\eps(X^j,r)\rho_\eps(r)\,drds)\,
\eeas
} \hfill \parbox{10mm}{\bea \label{guX}\eea}\\
The first term of this expression can be written as
\[
\su (  \de_i^1 \!  \underbrace{\intl_{-\eps}^u           \pa_i
f(x^j_\eps(X^k,s))\rho_\eps(r)\,dr}_{\to \pa_i f(X^j)\theta}
+ \underbrace{\frac{1}{2} \intl_{-\eps}^u           \pa_i
f(x^j_\eps(X^k,r))\rho_\eps(r)dr\,
\pa_1\!\!\intl_{-\eps}^u\intl_{-\eps}^s
 \pa_i f(x^j_\eps(X^k,r)\rho_\eps(r)\,drds}_{=:A})
\]
For any test function $\vphi$ we have
\[
\langle 2A, \vphi \rangle = \intl_{-\eps}^\infty \vphi(u)
\intl_{-\eps}^u           \pa_i
f(x^j_\eps(X^k,r))\rho_\eps(r)dr\,
\pa_1\!\!\intl_{-\eps}^u\intl_{-\eps}^s
 \pa_i f(x^j_\eps(X^k,r)\rho_\eps(r)\,drdsdu
\]
Splitting this integral into a sum of the form 
\beq \label{scheme}
\intl_{-\eps}^\eps \dots \intl_{-\eps}^u\dots \intl_{-\eps}^u
\intl_{-\eps}^s\dots\, +
\intl_{\eps}^\infty          \dots         \intl_{-\eps}^u\dots
\intl_{-\eps}^\eps \intl_{-\eps}^s \dots\,+
\intl_{\eps}^\infty          \dots         \intl_{-\eps}^u\dots
\intl_{\eps}^u \intl_{-\eps}^s \dots \,
\eeq
the  boundedness arguments following (\ref{dxX}) imply that the
first   two   summands  converge  to  $0$.  Also,  by  (11)  in
\cite{geo} and by (\ref{geo}) we obtain
\beq \label{Dif}
\lim_{\eps \to 0} \sup_{|r|<\eps} |\pa_i(f(x_\eps^j(X^k,r))) - 
\pa_i f(X^k) | = 0
\eeq
uniformly for $X^k$ in compact sets.
For later use we note that by the same reasoning also
\beq \label{Diff}
\lim_{\eps \to 0} \sup_{|r|<\eps} |\pa_i(\pa_j f(x_\eps^j(X^k,r))) - 
\pa_{ij} f(X^k) | = 0
\eeq
uniformly for $X^k$ in compact sets.
Hence by a direct estimation the limit of the remaining 
term is $\langle \pa_i f (X^j) \pa_1 \pa_i f (X^j) u_+, \vphi 
\rangle $, so the distributional limit of the first term in
(\ref{guX}) is $\pa_1 f (X^j) \theta + \frac{1}{2} \su 
\pa_i f (X^j) \pa_i \pa_1 f (X^j) u_+ $.

Turning  now  to the second term in (\ref{guX}) we obtain from
(\ref{Dif}) and (\ref{gendelta}b):
\[
\pa_1 \!\intl_{-\eps}^uf(x^j_\eps(X^k,r))\rho_{\eps}(r)\,dr \to
\pa_1 f (X^j) \theta.
\]
Inserting (\ref{geo}) it remains to calculate the distributional
limit of\\
\parbox{15cm}{
\beas 
&& \frac{1}{2}\intl_{-\eps}^u\intl_{-\eps}^s \pa_1(\pa_i 
 f(x^j_\eps(X^k,r))
 \rho_\eps(r) \intl_{-\eps}^r 
 \pa_i f(x^j_\eps(X^k,q) \rho_\eps(q)\,dq\,dr\,ds \\
&& +\frac{1}{2}\intl_{-\eps}^u\intl_{-\eps}^s \pa_i 
 f(x^j_\eps(X^k,r)
 \rho_\eps(r) \intl_{-\eps}^r 
 \pa_1(\pa_i f(x^j_\eps(X^k,q)) \rho_\eps(q)\,dq\,dr\,ds \,.
\eeas
} \hfill \parbox{10mm}{\bea \label{1}\eea}\\
A  splitting  scheme as in (\ref{scheme}) for both these
terms, (\ref{Dif}) and the fact that 
\[
\intl_{\eps}^u\intl_{-\eps}^s \rho_\eps(r)\intl_{-\eps}^r
\rho_\eps(q) dq\, dr\, ds \to \frac{1}{2} u_+
\]
imply that    each    summand   in   (\ref{1})   converges
distributionally  to  $\frac{1}{4}  \pa_i  f(X^j)\pa_{1i} f(X^j)
u_+$. 

An analogous argument holds for $g_{uY}$. Summing up, we obtain
\beq \label{guX0}
g_{uX} \stackrel{\DD'}{\longrightarrow} 0  \hspace{1cm}
\mbox{and} \hspace{1cm}
g_{uY} \stackrel{\DD'}{\longrightarrow} 0  \hspace{1cm} 
\mbox{for }\eps
\to 0 \,.
\eeq
If we write
\beq
g_{XX} = \su (\pa_1(X^i + \frac{1}{2} \intl_{-\eps}^u\intl_{-\eps}^s\pa_i
       	  f(x^j_\eps(X^k,r))\rho_\eps(r)\,dr\,ds))^2 
       =: \su R_i
\eeq
then
\[
R_i = \delta^1_i + \underbrace{
\intl_{-\eps}^u\intl_{-\eps}^s \pa_1(\pa_1
       	  f(x^j_\eps(X^k,r)))\rho_\eps(r)\,dr\,ds}_{\to  
       	  \pa_{11} f(X^j) u_+} + \underbrace{\frac{1}{4} (
	  \intl_{-\eps}^u\intl_{-\eps}^s \pa_1(\pa_i
       	  f(x^j_\eps(X^k,r)))\rho_\eps(r)\,dr\,ds )^2}_{=:B}
\]
and  by  a  similar  argument  as  above  it  follows  that the
distributional limit of the derivative $\pa_u B$ is
$\frac{1}{2}(\pa_1\pa_i  f(X^j))^2 u_+$.  Since  taking primitives
(i.e.   convoluting   with   $\theta$)  is  separately continuous  
on  the convolution algebra of distributions supported in an acute 
cone  and since taking tensor products of distributions is separately
continuous as well we get
\[
R_i \to \delta^1_i + \pa_{11}  f(X^j) u_+ 
+ \frac{1}{2}(\pa_1\pa_i  f(X^j))^2  \frac{u_+^2}{2}\,.  
\]
Thus
\bea
&& g_{XX} \stackrel{\DD'}{\longrightarrow} 
(1 + \frac{1}{2}\pa_{11} f(X^j)u_+)^2
   + \frac{1}{4}(\pa_{12} f(X^j))^2 u_+^2 \label{gXX}\\
&& g_{YY} \stackrel{\DD'}{\longrightarrow} 
(1 + \frac{1}{2}\pa_{22} f(X^j)u_+)^2
   + \frac{1}{4}(\pa_{12} f(X^j))^2 u_+^2 \label{gYY}
\eea
Finally, we turn to $g_{XY} = 2\su (\pa_1 x^i_\eps(X^j))
(\pa_2 x^i_\eps(X^j)) =: 2\su S_i$. Inserting from (\ref{geo})
we have 
\beas
&& S_i = \frac{1}{2}\underbrace{
\intl_{-\eps}^u\intl_{-\eps}^s \pa_2(\pa_1
       	  f(x^j_\eps(X^k,r)))\rho_\eps(r)\,dr\,ds}_{\to  
       	  \pa_{12}f(X^j) u_+} + 
\frac{1}{2}\underbrace{
\intl_{-\eps}^u\intl_{-\eps}^s \pa_1(\pa_2
       	  f(x^j_\eps(X^k,r)))\rho_\eps(r)\,dr\,ds}_{\to  
       	  \pa_{12}f(X^j) u_+} + \\	  
&&	  \frac{1}{4}\underbrace{ 
	  \intl_{-\eps}^u\intl_{-\eps}^s \pa_2(\pa_i
       	  f(x^j_\eps(X^k,r)))\rho_\eps(r)\,dr\,ds
	  \intl_{-\eps}^u\intl_{-\eps}^s \pa_1(\pa_i
       	  f(x^j_\eps(X^k,r)))\rho_\eps(r)\,dr\,ds}_{=:D}
\eeas
We claim that 
\beq \label{claim}
D \to \pa_{1i} f(X^j) \pa_{2i} f(X^j) u_+^2.
\eeq
To establish this, let $\vphi \in \DD$ and consider 
\beas
&&\langle D,\vphi \rangle - \pa_{1i} f(X^j) \pa_{2i} 
f(X^j) \intl_0^\infty u^2 \vphi(u) \,du = 
\intl_{-\eps}^\infty \left(\intl_{-\eps}^u\intl_{-\eps}^s \pa_1(\pa_i
       	  f(x^j_\eps(X^k,r)))\rho_\eps(r)\,dr\,ds\right)\\
&&	  \left(\intl_{-\eps}^u\intl_{-\eps}^s \pa_2(\pa_i
       	  f(x^j_\eps(X^k,r)))\rho_\eps(r)\,dr\,ds\right)
       	  \vphi(u)\,du - 
\intl_0^\infty \pa_{1i} f(X^j) \pa_{2i} 
f(X^j)  u^2 \vphi(u) \,du
\eeas
In  order to show that this goes to $0$, by a splitting similar
to  (\ref{scheme})  and  by  the boundedness properties already
established it suffices to prove
\beas
&& 
\intl_{\eps}^\infty \left(\intl_{\eps}^u\intl_{-\eps}^s \pa_1(\pa_i
       	  f(x^j_\eps(X^k,r)))\rho_\eps(r)\,dr\,ds\right)
	  \!\! \left(\intl_{\eps}^u\intl_{-\eps}^s \pa_2(\pa_i
       	  f(x^j_\eps(X^k,r)))\rho_\eps(r)\,dr\,ds\right)
       	  \vphi(u)\,du - \\
&& - \intl_\eps^\infty \pa_{1i} f(X^j) \pa_{2i} 
f(X^j)  u^2 \vphi(u) \,du  \to 0\, ,
\eeas
which   in   turn   is   a   consequence  of  (\ref{Diff})  and
(\ref{gendelta}). Therefore,
\beq \label{gXY}
g_{XY} \stackrel{\DD'}{\longrightarrow} \frac{1}{2} u_+^2
\left(\pa_{11}f(X^j)\pa_{21}f(X^j) + 
\pa_{12}f(X^j)\pa_{22}f(X^j)\right)
+ 2 u_+ \pa_{12}f(X^j)
\eeq From  
(\ref{guX0}), (\ref{gXX}), (\ref{gYY}) and (\ref{gXY}) we
obtain  the  distributional  limit of the regularized metric in
the form
\bblock
&& ds_\eps^2 \stackrel{\DD'}{\longrightarrow} -du dV +
(1 + \frac{1}{2} \pa_{11}f(X^j)u_+)^2 dX^2 +
(1 + \frac{1}{2} \pa_{22}f(X^j)u_+)^2 dY^2 +\\
&& 
\hphantom{ds_\eps^2 \stackrel{\DD'}{\longrightarrow}}
+ \frac{1}{2} \pa_{12}f(X^j)\triangle f(X^j) u_+^2 dX dY
+ 2 u_+ \pa_{12}f(X^j) dX dY+\\
&&
\hphantom{ds_\eps^2 \stackrel{\DD'}{\longrightarrow}}
+\frac{1}{4} (\pa_{12}f(X^j))^2 u_+^2 (dX^2\! +\! dY^2)\,.
\eblock{dslim}
Let us sum up the main result of this section in the following
diagram \\
\setlength{\unitlength}{1cm}
\begin{picture}(15,6)
\put(0.3,5){$ds^2=f(x^i)\de(u)du^2-dudv+\sum (dx^i)^2$}
\put(9.5,5){$ds_\eps^2=f(x^i)\rho_\eps(u)du^2-dudv+\sum (dx^i)^2$}
\put(7,5){\vector(1,0){2}}
\put(7.7,5.3){reg.}
\put(13,4.5){\vector(0,-1){2.5}}
\put(3,4.5){\vector(0,-1){2.5}}
\put(3.2,3.2){??}
\put(10,1){
\parbox{6cm}{$ds_\eps^2=
	-dudV+(2\sum\limits_{i=1}^2(\dot x^i_\eps
	\pa_jx^i_\eps)-\pa_jv_\eps)dudX^j
	+\sum\limits_{i=1}^2(\pa_jx^i_\eps dX^j)^2$}}
\put(1.8,1){(\ref{dslim}) = $\lim\limits_{\eps\to 0} ds_\eps^2$}
\put(9,1){\vector(-1,0){2}}
\put(7.5,1.3){$\mathcal D'$-limit}
\put(13.5,3.2){$T_\eps$}
\end{picture}\\
The left hand arrow in the above diagram represents the pullback  
of  the  distributional  metric (\ref{metric}) under the
discontinuous coordinate change (see (\ref{res1}))\\
\parbox{15cm}{
\beas
 x^i &=& X^i + \frac{1}{2} \pa_i f(X^j) u_+\\
 v    &=&  V  + f(X^j)\theta(u) + \frac{1}{4} \pa_i f(X^j)\pa^i
 f(X^j) u_+ \,.
\eeas
} \hfill \parbox{10mm}{\bea \label{dct}\eea}\\
Note that (\ref{dct}) is the generalization of the Penrose junction
conditions (\ref{jc}) to the case of an arbitrary wave profile $f$
(cf. (16) in \cite{aib}).
Although  undefined  within  Schwartz distribution theory, this
transformation can be interpreted consistently as the  composition 
of a regularization  procedure, a  smooth transformation of  
the    regularized   metric  and  a distributional limit. It is
precisely      in     this   sense   that  the  pullback 
is given by the right hand side of (\ref{dslim}) or for short
\beq\label{disttrans}	
	ds^2\,=\,-du dV+(\de_{ij}+\frac{1}{2}u_+\pa_{ij}f)\,
	(\de^j_k+\frac{1}{2}u_+\pa^j_kf)\, dX^i dX^k
\eeq
For  the  special  case  of an impulsive plane wave
$f(x,y) = x^2 - y^2$ this exactly reduces to (\ref{cm}) and in the general 
vacuum case it is equivalent to (2) in~\cite{pv1}. Also, we note that 
(\ref{disttrans}) appeared first as (17) in \cite{aib}.
The   regularization procedure  employed above justifies {\em a
posteriori} the derivation of (\ref{disttrans}) by using formal
multiplication rules for distributions 
(cf. the remarks following (\ref{jc})).
\section{Generalized Coordinate Transformations}
In  this  section we are going to give a precise meaning to the
left  hand  arrow  in the above diagram in terms of Colombeau's
generalized  functions.  In  order  to facilitate the following
argument  we  are  going to work in the so called  `simplified'
algebra  (cf. \cite{geo2} for a treatment of impulsive
pp-waves in the full Colombeau algebra). We note, however, that
the  considerations  to  follow  retain their validity in the 
`full' setting as well.
Let   $\Omega$  be  an  open  subset  of  $\R^n$.  With  ${\cal
E}(\Omega;\R^n)
= ({\cal C}^\infty (\Omega;\R^n))^{(0,\infty)}$ set
\begin{eqnarray*} 
&&{\cal E}_M(\Omega;\R^n):=\{(u_\eps)_\eps\in 
{\cal E}(\Omega;\R^n): \forall K\subset\subset\Omega, \forall m\in\N_0
\, \exists p\in \N \,:\,\\
&& \hphantom{{\cal E}_M(\Omega;\R^n) :=\,\,}
\sup_{x\in K}\|D^{(m)} u_\eps
(x)\|=O(\eps^{-p})\, (\eps\rightarrow 0)\}\\
&&{\cal N}(\Omega;\R^n):=\{(u_\eps)_\eps\in 
{\cal E}(\Omega;\R^n): 
\forall K\subset\subset\Omega, \forall m\in\N_0
\mbox{ }\forall q\in\N \,:\, \\
&& \hphantom{{\cal N}(\Omega;\R^n):=\,\,}
\sup_{x\in K}\|D^{(m)} u_\eps
(x)\|=O(\eps^{q})\,(\eps\rightarrow 0)\}
\end{eqnarray*}
where $D^{(m)}$ denotes the $m$-th derivative and  
$\|\,.\, \|$ is any norm on ${\cal L}((\R^n)^m;\R^n)$.
Then  the  simplified  Colombeau  algebra ${\cal G}(\Omega;\R^n)$ is
defined      as      the  differential  factor  algebra  ${\cal
E}_M(\Omega;\R^n)/{\cal
N}(\Omega;\R^n)$ (see \cite{mo,c1,AB}).

Now    let  $\Omega$, $\Om'$  be  open  subsets of
$\R^4$ whose $u$-projection
contain   the  interval  $[-1,0]$.  It  is  a  straightforward
consequence  of Theorem 1 in \cite{geo2} and the remarks following 
(\ref{dxX})  that  $(T_\eps)_\eps$ and $(T_\eps^{-1})_\eps$ (if
defined on $\Om'$) are
elements         of    ${\cal    E}_M(\Omega;\R^4)$  and  ${\cal  
E}_M(\Omega';\R^4)$, respectively.  Let  us  denote  their classes
in ${\cal G}(\Omega;\R^4)$ (resp. ${\cal G}(\Om';\R^4)$) by $T$ and $S$.
We claim that for a suitable choice of $\Om$, $\Om'$, $S$ is the 
inverse in $\cal G$ of $T$, which will
justify viewing $T$ as a generalized coordinate transformation.

Although     composition  of  Colombeau  generalized  functions
(through componentwise insertion of representatives) is not
always  well defined, a sufficient condition for this operation
to        yield   a    unique  Colombeau  function was given in
\cite{AB}, 7.3.1 and 7.4.1, to
wit:    Let    $U\in    {\cal    G}(\Om';\R^n)$,    $V\in {\cal
G}(\Om;\R^n)$, then
$U\circ  V  :=  [(U_\eps\circ  V_\eps)_\eps]$ is a well defined
element  of  ${\cal  G}(\Om;\R^n)$  if there exists a representative
$(U_\eps)_\eps$ of $U$ such that
\beq \label{gcomp}
\forall K\subset\subset  \Om  \,  \exists \, K'  \subset\subset \Om' \,
\, \exists \eta>0 \,\, {\mathrm s.t.} \,\, U_\eps(K) \subseteq K'
\, \, \forall \eps < \eta \,.
\eeq
It  follows  immediately from the remarks following (\ref{dxX})
that  $S$ satisfies (\ref{gcomp}). Concerning $T$, suppose that
$(u,v,x,y)$ varies in some $K\subset\subset \Omega$ and set
$(u,V_\eps,X_\eps,Y_\eps) = T_\eps(u,v,x,y)$. Then 
\[
x^i = x^i_\eps(X^j_\eps,u)= X^i_\eps + \frac{1}{2} 
\intl_{-\eps}^u\intl_{-\eps}^s \pa_i  f  (x^j_\eps(X^k_\eps,r))
\rho_\eps(r)\,dr\,ds\,.
\] 
Since the left hand side and the argument of $\pa_i f$ in this 
equation are bounded by assumption, $X^i_\eps$ remains in some 
compact region $L$  in  $\R^2$.
We already know that on any
such  set  $L$  (times  some  compact  $u$-interval) $\dot
x^i_\eps$ is uniformly bounded. Inserting this into
\[
V_\eps =
v_\eps(V_\eps,X^i_\eps,u) - \intl_{-\eps}^u f(x^j_\eps(X_\eps^k,s))
  \rho_\eps(s)\,ds - \intl_{-\eps}^u\intl_{-\eps}^s\pa_i f(
 x^j_\eps(X^k_\eps,r))\dot x^i_\eps(X_\eps^k,r))\rho_\eps(r)\,dr\,ds
\]
also establishes the desired boundedness property for $V_\eps$.
We conclude that  $S$  is  indeed the $\cal
G$-inverse        of        $T$ (for suitable regions $\Omega$,
$\Omega'$). Therefore the left hand arrow in the above
diagram, i.e. the (generalized) Penrose junction conditions (\ref{dct})
turn out
to be the macroscopic aspect (or `distributional 
shadow')  of the (pullback under the) generalized coordinate 
transformation $T$. 
\section{Conclusions}\label{conclusions}
In this work we have studied the discontinuous coordinate transformation
relating the distributional with the continuous form of the impulsive pp-wave 
metric. It is indeed possible to interpret this change of coordinates as the
(distributional) limit of a family of smooth transformations which we 
obtained by a {\em general} regularization proceedure. We emphasize the fact
that our results are completely regularization independent even within
the maximal class of regularizations of the Dirac-$\de$, hence ``natural''.
Note also that we did not have to impose Einstein's equations throughout our
analysis, hence our results apply to all pp-waves, i.e. even to the 
non-vacuum case.
  
Physically speaking our regularization approach consists in viewing the
impulsive wave as a limiting case of a sandwich wave with an arbitrarily
regularized wave profile. From this point of view the two forms of the
impulsive  metric  arise  as the (distributional) limits of the
sandwich wave in different coordinate systems.
It is a well known fact~\cite{geroch} that given a one-parameter
family of spacetimes one may obtain different limits by introducing a
suitable (parameter dependent) change of coordinates. This effect
is mainly caused by a different identification of points in the manifold.
More precisely, the transformation in the limit sends finite points
to infinity, hence is unbounded. However, in our case the limit     
of the transformation, although discontinuous, remains bounded and the
mentioned ambiguities do not arise. This is in fact not surprising since
our change of coordinates is adjusted to the geometry of the spacetime
which (in the sense of geodesic completeness) is nonsingular,
allowing only for a {\em finite} jump of the geodesics.

Finally, the family of
smoothened transformations constitutes a generalized coordinate
transformation in the sense of Colombeau, whose macroscopic (i.e. 
distributional) aspect is precisely given by the (generalized) Penrose 
junction conditions (\ref{dct}). 

These results  put the formal
calculations  (using  certain  ad-hoc  multiplication rules for
distributions) on solid ground and set up a framework which allows to 
also {\em mathematically}  identify  the two ``physically equivalent'' 
descriptions of impulsive waves.
\section*{Acknowledgement}
It's a pleasure for us to thank all the members of the DIANA research group 
{\sf (http://diana.mat.univie.ac.at/\~\,diana) } for providing such an inspiring
working environment. We also thank H. Urbantke for his comments
on the draft version of this paper.
M. Kunzinger was supported by Research Grant P10472-MAT of the Austrian 
Science Foundation. R. Steinbauer was supported by Austrian Academy of 
Science, Ph.D. programme, grant \#338 and by Research Grant P12023-MAT 
of the Austrian Science Foundation.
                 
\end{document}